\documentclass[conference]{IEEEtran}
\IEEEoverridecommandlockouts

\usepackage{cite}
\usepackage{amsmath,amssymb,amsfonts}
\usepackage{algorithmic}
\usepackage{graphicx}
\usepackage{textcomp}
\usepackage{xcolor}

\usepackage{booktabs}
\usepackage{caption}
\usepackage{array}
\usepackage{multirow}
\usepackage{arydshln}
\usepackage{threeparttable}
\usepackage[normalem]{ulem}  

\def\BibTeX{{\rm B\kern-.05em{\sc i\kern-.025em b}\kern-.08em
    T\kern-.1667em\lower.7ex\hbox{E}\kern-.125emX}}
\begin{document}

\title{Ges-QA: A Multidimensional Quality Assessment Dataset for Audio-to-3D Gesture Generation\\

}

\author{\IEEEauthorblockN{Zhilin Gao, Yunhao Li, Sijing Wu, Yuqin Cao, Huiyu Duan, Guangtao Zhai}
\IEEEauthorblockA{\textit{Shanghai Jiao Tong University, Shanghai, China} \\
\{undefined49527, lyhsjtu, wusijing, caoyuqin, huiyuduan, zhaiguangtao\}@sjtu.edu.cn
}
}

\maketitle

\begin{abstract}

The Audio-to-3D-Gesture (A2G) task has enormous potential for various applications in virtual reality and computer graphics, etc. However, current evaluation metrics, such as Fréchet Gesture Distance or Beat Constancy, fail at reflecting the human preference of the generated 3D gestures. To cope with this problem, exploring human preference and an objective quality assessment metric for AI-generated 3D human gestures is becoming increasingly significant. In this paper, we introduce the Ges-QA dataset, which includes 1,400 samples with multidimensional scores for gesture quality and audio-gesture consistency. Moreover, we collect binary classification labels to determine whether the generated gestures match the emotions of the audio. Equipped with our Ges-QA dataset, we propose a multi-modal transformer-based neural network with 3 branches for video, audio and 3D skeleton modalities, which can score A2G contents in multiple dimensions. Comparative experimental results and ablation studies demonstrate that Ges-QAer yields state-of-the-art performance on our dataset.

\end{abstract}


\section{Introduction}

The Audio-to-3D-Gesture (A2G) task has attracted a lot of attention due to its potential for the development of digital human generation \cite{beat} and embodied agents. Many researchers are dedicated to developing methods to generate high-quality synchronized 3D gestures driven by speech. A subset of approaches concentrates on refining facial expressiveness \cite{wu2024mmhead, wu2023ganhead, wu2023singinghead}, while others investigate full-body avatar generation \cite{gestureLSM}. Techniques such as VQ-VAE \cite{probtalk, talkshow} and diffusion models \cite{amuse, syntalker, diffsheg} have demonstrated state-of-the-art performance. Some methods also attempt to combine more modalities, which can accept text or other types of input together with audio \cite{lom, motioncraft, li2025samr}. Although these approaches have made progress in various aspects, the evaluation of quality of generated 3D gestures still faces challenges. For example, the commonly used Frechet Gesture Distance (FGD) only computes the distributional similarity between the ground truth gestures and generated gestures. Beat Constancy only uses the amplitude of changes in joints and audio to measure the audio-visual consistency \cite{emage}. These metrics cannot comprehensively evaluate the human preference on generated gestures. Hence, a subjective and objective quality assessment experiment is crucial.

The quality assessment of 3D Dynamic Digital Humans (DDHs) primarily focuses on distortion in the AI generation process \cite{zhang2024human, li2025aghi}. For A2G tasks, certain approaches evaluate exclusively on head region \cite{wu2025fvq, yang2025lmme3dhf}, while others incorporate reference videos to enhance evaluation robustness \cite{6faceFR, 27wholeRR}. As DDHs are typically presented through 2D-rendered animation videos, audio-visual quality assessment methodologies can be naturally extended to related evaluation tasks \cite{25wholeDynamic, 26wholeNR}. With the rapid development of deep learning, researchers have more tools to utilize, such as alignment methods \cite{avid_cma,valor,vast} and deep learning-based methods \cite{anna2,generalAVQA2}. However, compared to conventional AIGC, the distinguishing characteristics of A2G content remain substantially underexplored in recent quality assessment research, including unnatural limb movements across temporally adjacent frames and speech-action desynchronization.

\begin{figure}[t]
\centering
\includegraphics[width=0.45\textwidth]{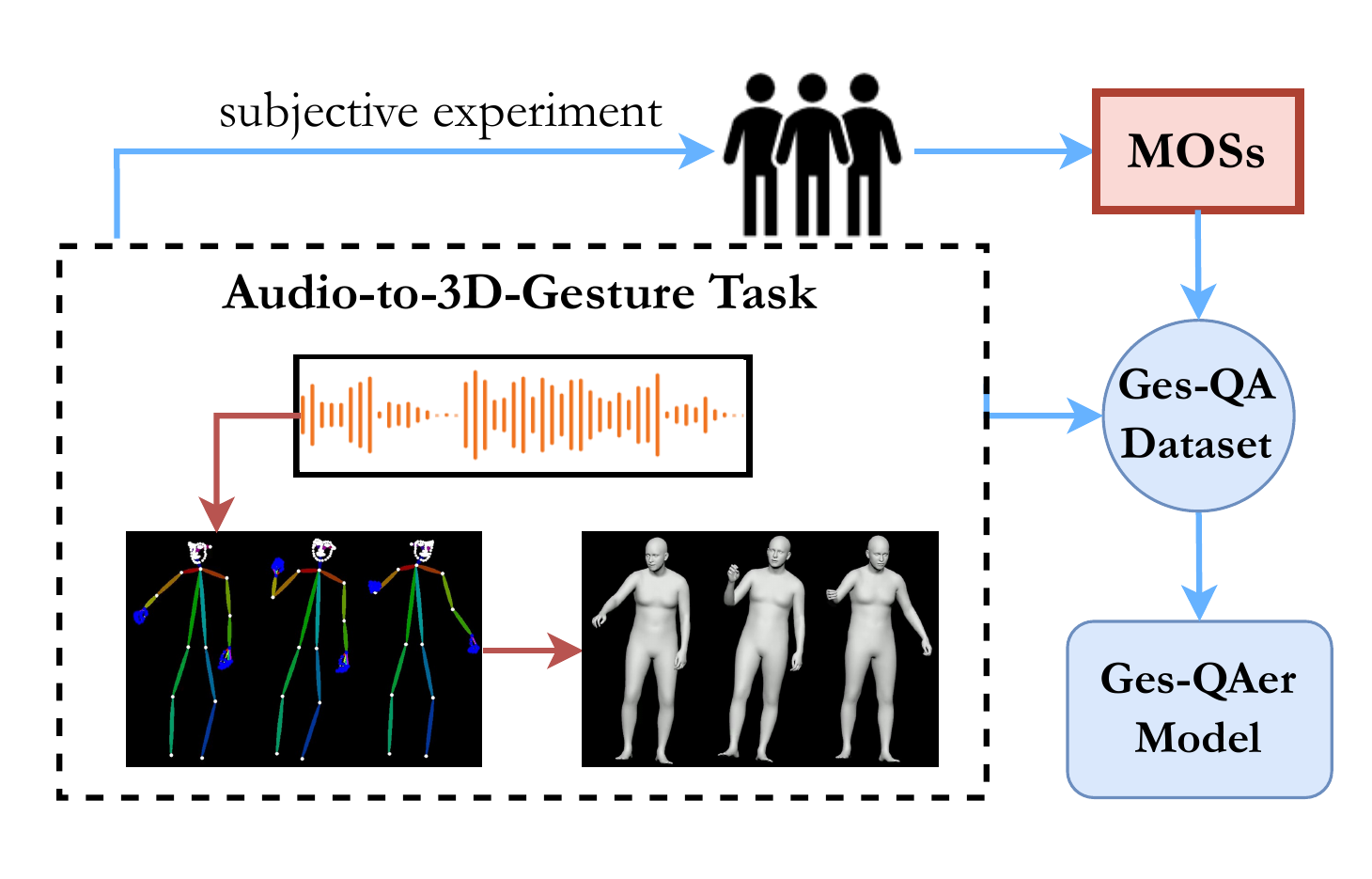}
\caption{The pipeline overview of A2G quality assessment task. Following data acquisition and Ges-QA dataset establishment via subjective experiment, we train an operational quality evaluation framework Ges-QAer.}
\label{fig1}
\vskip -0.2in
\end{figure}

\begin{figure}[t]
\centering
\includegraphics[width=\columnwidth]{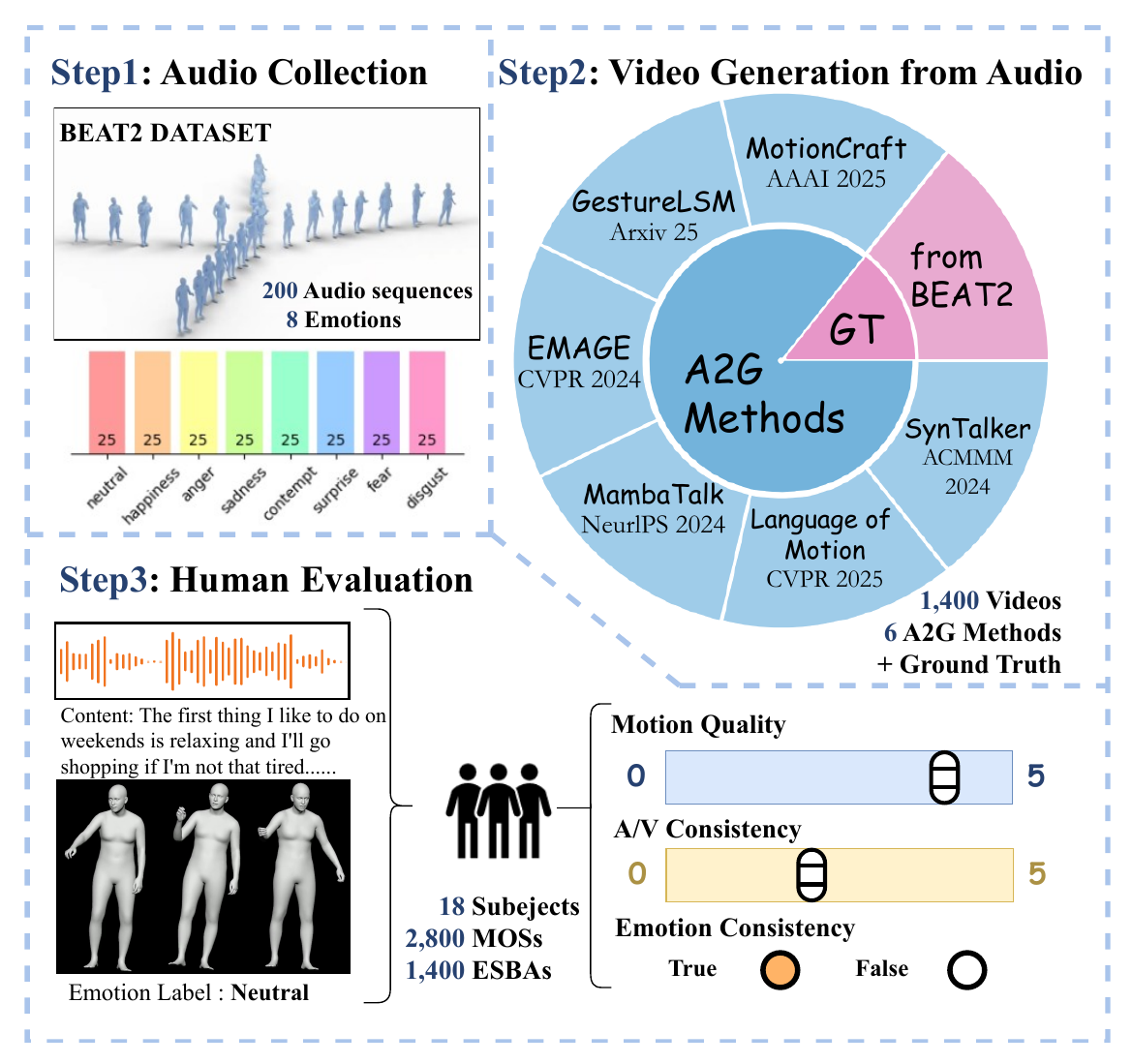}
\caption{The construction process of Ges-QA dataset. Step 1 displays eight emotion categories of audio data. In Step 3, subjects will provide two Mean Opinion Scores (MOSs) from different dimensions and one Subjective Binary Annotation for Emotion (ESBA) for each sample.}
\label{fig2}
\vskip -0.15in
\end{figure}

To solve this problem, we constructed the first A2G quality assessment dataset, \textbf{Ges-QA}, which includes 1,400 A2G samples generated by 6 approaches and ground truth (GT) data. Fig.~\ref{fig2} shows the process of constructing the dataset. After obtaining the generated 3D gestures, we carefully conduct the subjective experiment to collect the Mean Opinion Scores (MOSs) from the dimensions of gesture quality and audio-gesture consistency. Notably, to investigate the ability of current A2G approaches to handle emotions, we simultaneously collected subjects' opinions on the emotion congruence status of A2G contents. We evaluated the performance of existing AVQA methods on the Ges-QA dataset, as shown in Tab.~\ref{tab1}. The results indicate that the AVQA methods still have considerable potential for refinements in evaluating the quality of A2G contents.

We further propose the first quality assessment method for A2G, \textbf{Ges-QAer}. As shown in Fig.~\ref{fig5}, Ges-QAer encodes vision, audio and motion separately with three single-modality encoders, and outputs predicted numerical scores for A2G contents. Specifically, the Ges-QAer model projects three modalities into the same common space and learn how to predict the multidimensional quality scores for A2G contents. Ablation studies have been conducted to demonstrate the effectiveness of the proposed Ges-QAer model.

Our experimental results indicate that Ges-QAer achieves state-of-the-art performance on the Ges-QA dataset. Our core contributions can be summarized into two points:
\begin{itemize}
\item \textbf{A large-scale quality assessment dataset for A2G task Ges-QA.} It annotates A2G quality with two-dimensional scores: gesture quality and audio-gesture consistency. And we tentatively studied the problem of emotion congruence in the A2G task.
\item \textbf{A novel quality assessment model, Ges-QAer.} It can predict multidimensional quality scores for A2G content, offering users a better audio-visual experience.  
\end{itemize}

\section{Dataset Construction}

Our proposed Ges-QA dataset is designed for multidimensional score prediction. In this section, we introduce the construction process, as illustrated in Fig.~\ref{fig2}, and analyze the subjective scores. As a supplement, we evaluated the ability of the A2G approaches involved in emotional processing through subjects' opinions on whether the generated gestures match the emotions of the speech.

\subsection{Data Generation}

\paragraph{Audio Collection} 
We selected BEAT2 \cite{beat,emage} as the audio data source. BEAT2 is a holistic and high-quality 3D motion captured dataset, consisting of 60 hours of data for 25 speakers (in English). For each speaker, data are recorded with eight emotions. We selected 25 sequences of 10 seconds for each emotion. To ensure the diversity of data, only four recordings come from the same two speakers. Step 1 of the dataset construction process in Fig.~\ref{fig2} shows the eight emotions.

\paragraph{Motion Generation from Audio} 
We utilized six latest A2G approaches, including EMAGE \cite{emage}, MambaTalk \cite{mambatalk}, Syntalker \cite{syntalker}, Language of Motion (LoM) \cite{lom}, MontionCraft \cite{motioncraft}, GestureLSM \cite{gestureLSM}, to generate videos samples from audio using their default weights and code. Additionally, we also used the ground truth (GT) motion data provided by BEAT2. In the end, we obtained a total of 1,400 A2G samples ((6 approaches + GT) $\times$ 200 audio). Notably, 3D skeleton data were retained after video rendering because they can serve as training input for the quality assessment model.

\paragraph{Human Evaluation}
We invited 18 subjects to participate in our subjective experiment. Subjects were asked to rate A2G samples across two dimensions: gesture quality and audio-gesture consistency. Gesture quality assesses the perceived quality of motions, including naturalness and similarity to the real world. Audio-gesture consistency mainly evaluates whether the motion is consistent with the rhythm of the speech. Equally important, to investigate the ability of current A2G methods to handle emotions, we simultaneously collected subjects' opinions on the emotion congruence status of A2G contents. Subjects will give a binary judgment to evaluate whether gestures and speech express the same emotion. This judgment is referred to as Subject Binary Annotation for Emotion (ESBA). These data were collected via our custom interface featuring two Likert scale sliders and a radio button for ESBA selection. Likert scores were normalized to Z-scores ranging from 0 to 100, with Mean Opinion Scores (MOSs) derived from averaged Z-scores. ESBA values were determined through majority voting after outlier exclusion.

\begin{figure}[hb]
\centering
\includegraphics[width=0.5\textwidth]{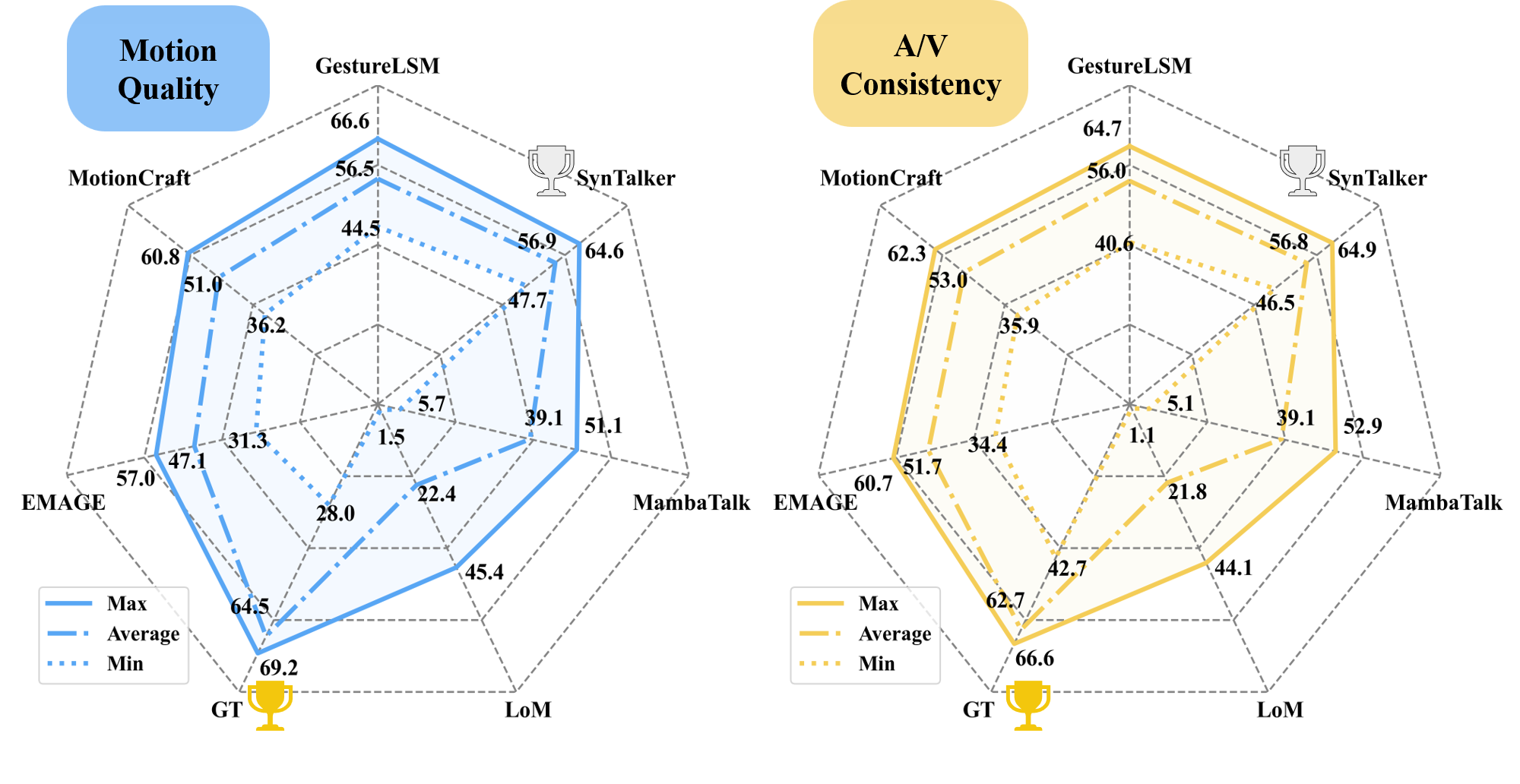}
\caption{Mean Opinion Score comparison of gesture quality and audio-gesture consistency across multiple A2G approaches.}
\label{fig3}
\end{figure}

\subsection{MOSs Analysis}
Fig.~\ref{fig3} illustrates the maximum, minimum, and average subjective scores of six A2G approaches (and GT data) across the two dimensions. We can observe that SynTalker exhibits quality second only to ground truth data, attributed to its emphasis on the elaborate control of synergistic full-body motion. Some approaches perform poorly, perhaps due to their focus more on multi-modal collaborative work than single A2G task.

\begin{figure}[t]
\centering
\vskip -0.1in
\includegraphics[width=0.5\textwidth]{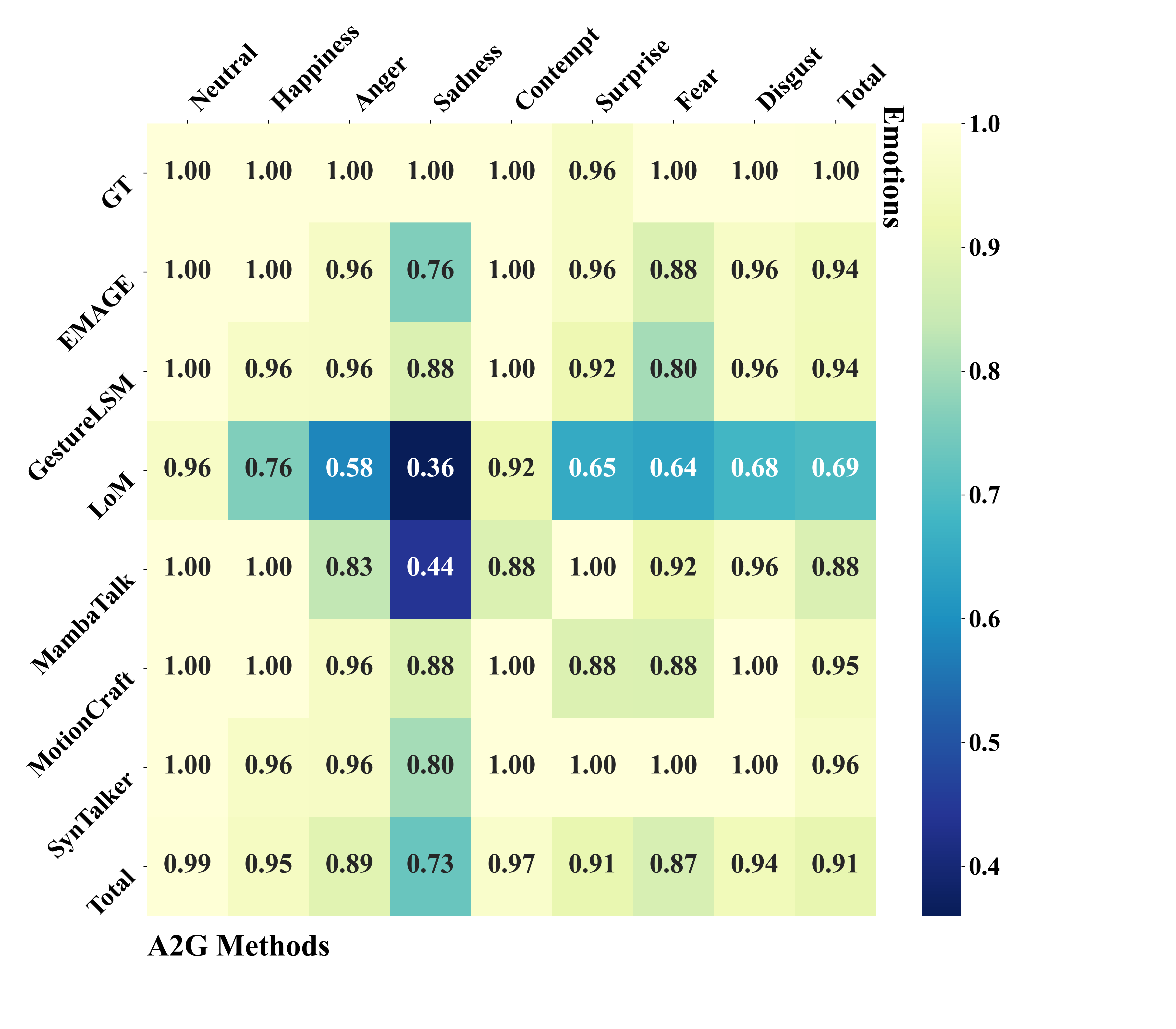}
\vskip -0.15in
\caption{Visualization of emotion congruence accuracy across multiple A2G approaches under different emotions.}
\label{fig4}
\vskip -0.15in
\end{figure}

\subsection{Emotional Annotation Analysis}

We quantified emotion congruence accuracy for each method through a statistical analysis of ESBAs. As seen in Fig.~\ref{fig4}, in most cases, the motion sequences provided by the A2G approaches can effectively restore the emotions contained in the speech input. A comparative analysis of Figures 3 and 4 revealed a strong positive correlation between emotion congruence accuracy and MOSs, tentatively suggesting that gesture quality modulates subjects' emotional perception.  However, intrinsic emotions such as "sadness" and "fear" are difficult to visually express through actions, so the performance of these A2G approaches is relatively poor. Emotions such as "happiness" and "contempt" are often better expressed through body language, resulting in better performance on ESBA.

\section{Ges-QAer Model}
\begin{figure}[hb]
\centering
\includegraphics[width=0.5\textwidth]{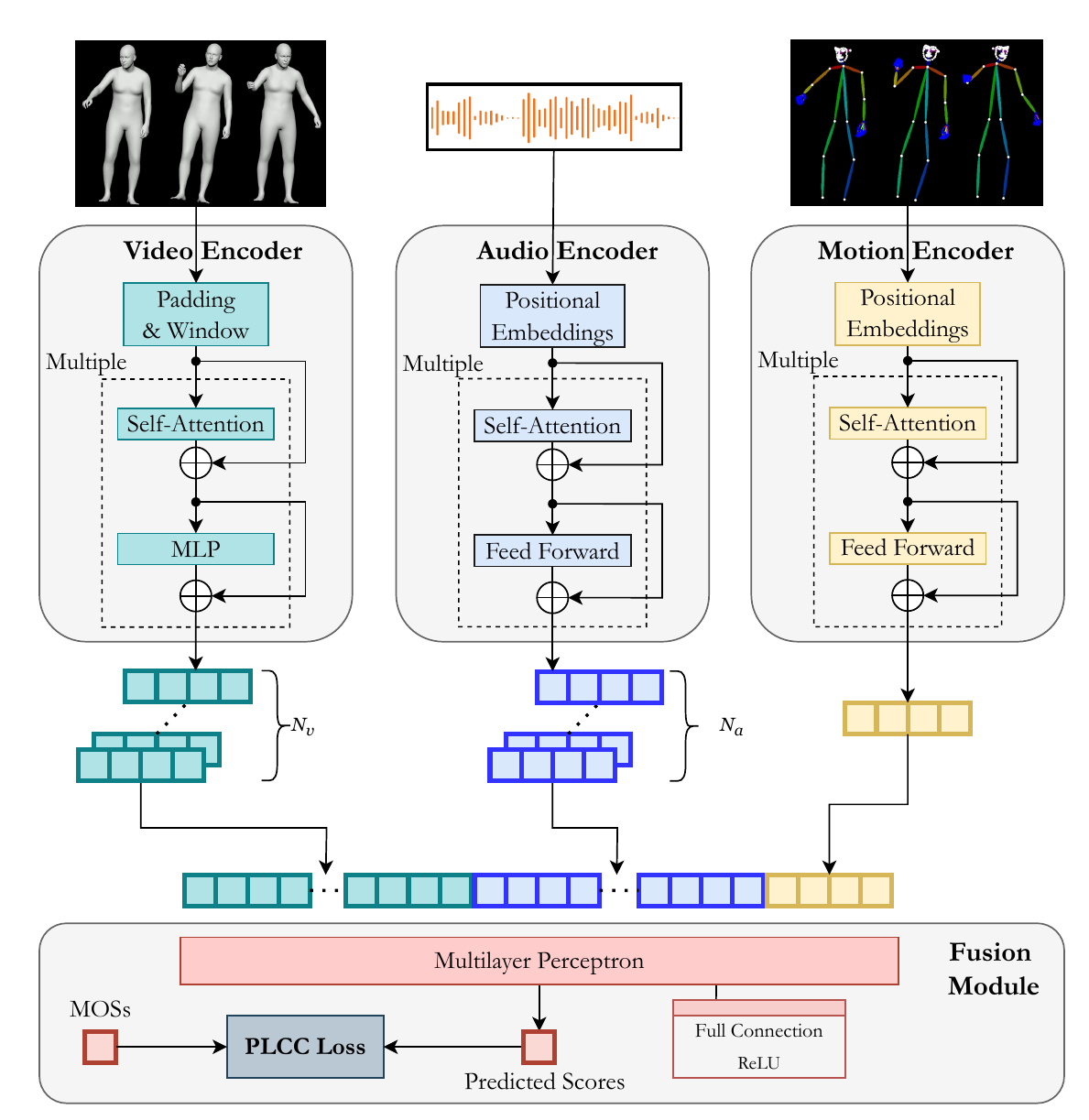}
\caption{The architecture of Ges-QAer. Ges-QAer uses three separate encoders to achieve single-modality representations, and a Multilayer Perceptron (MLP) for feature fusion.}
\label{fig5}
\end{figure}

We expect that Ges-QAer model can achieve end-to-end training without the need for pre- feature extraction, and possess specialized multi-modal capability for A2G task. To this end, we made dedicated designs about model architecture, as shown in Fig.~\ref{fig5}. We take the 3D skeleton information, which is usually the default output by the A2G task, as input along with the audio/video stream. After feature extraction and fusion, the model directly returns predicted numerical scores in both dimensions: gesture quality and audio-gesture consistency.

\subsection{Model Architecture}

Ges-QAer integrates three encoders for video, audio, and motion input. This architecture enables dedicated encoders to specialize in single-modality learning and facilitates pre-trained parameter inheritance, thus speeding up convergence and enhancing performances.

\paragraph{Vision Encoder}

Video Swin Transformer \cite{videoswin} is selected to process video inputs. For each video segment, $N_v$ frames are sampled and patched to produce features $F_v \in \mathbb{R}^{B \times N_v \times C} $, where $B$ is the batch size, and $C$ is the hidden dimension. These features are subsequently refined through transformer blocks equipped with 3D shifted-window based multi-head self-attention modules.

\paragraph{Audio Encoder}

We adopt Audio Spectrogram Transformer (AST) \cite{ast1,ast2} pre-trained on AudioSet as the audio encoder. AST is a convolutional-free and purely attention-based architecture. For each input stream, $N_a$ temporally segmented 5-second clips are transformed into spectrograms using a Hamming-windowed log-Mel filterbank. Through an embedding layer and transformer encoder modules, structured outputs $F_a \in \mathbb{R}^{B \times N_a \times C} $ are generated.

\begin{table*}[tbp]
\small
\centering
\renewcommand\arraystretch{1.03}
\renewcommand\tabcolsep{4pt}
\vskip -0.1in
\caption{Performance comparisons of our proposed Ges-QAer versus compared approaches and ablation variants on the Ges-QA dataset from two dimensions. Best results are bolded, second-best are underlined (ranking excludes ablation studies).}
\resizebox{\linewidth}{!}
{
\begin{tabular}{cl|cccc|cccc}
\hline

\multicolumn{2}{c|}{Dimension} & \multicolumn{4}{c|}{Gesture Quality} & \multicolumn{4}{c}{Audio-Gesture Consistency} \\
\hdashline
Model Type & Model & SRCC $\uparrow$ & PLCC $\uparrow$ & KRCC $\uparrow$ & RMSE $\downarrow$ & SRCC $\uparrow$ & PLCC $\uparrow$ & KRCC $\uparrow$ & RMSE $\downarrow$  \\
\hline
\multirow{3}{*}{\shortstack{Multi-modal \\ Alignment}}& AVID-CMA (CVPR 2021) & 0.2289 & 0.3528 & 0.1578 & 13.0465 & 0.0355 & 0.3297 & 0.0235 & 13.0856 \\
~ & VAST (NIPS 2023) & 0.5355 & 0.5260 & 0.3665 & 11.9890 & 0.1659 & 0.2511 & 0.1174 & 13.5459  \\
~ & ImageBind (CVPR 2023) & 0.4946 & 0.5212 & 0.3585 & 12.2693 & 0.2224 & 0.3113 & 0.3255 & 12.2821 \\
\hdashline
~ & DNN-RNT (TIP 2023) & 0.5607 & 0.6419 & 0.3982 & 9.7935 & 0.3746 & 0.5279 & 0.6783 & 10.9332  \\
AVQA & DNN-SND (TIP 2023) & 0.3025 & 0.3427 & 0.2077 & 12.5654 & 0.2138 & 0.3106 & 0.3927 & 12.3239 \\
~ & GeneralAVQA (TIP 2023) & \underline{0.8122} & \underline{0.8734} & \underline{0.6289} & \underline{6.8610} & \underline{0.7845} & \underline{0.8913} & \underline{0.6014} & \underline{6.3502} \\
\hline
\multicolumn{2}{c|}{\textbf{Ges-QAer} (Ours)} & \textbf{0.9282} & \textbf{0.9352} & \textbf{0.7687} & \textbf{4.9838} & \textbf{0.8795} & \textbf{0.9344} & \textbf{0.7070} & \textbf{4.9749}  \\
\hdashline
\multirow{2}{*}{Ablation} & \emph{w/o} video encoder & 0.0943 & 0.1229 & 0.0643 & 13.9742 & 0.0863 & 0.1278 & 0.0587 & 13.8898  \\
~ & \emph{w/o} motion encoder & 0.9280 & 0.9324 & 0.7698 & 5.0822 & 0.8760 & 0.9310 & 0.7015 & 5.1044  \\
\hline
\end{tabular}
}
\vskip -0.17in
\label{tab1}
\end{table*}

\paragraph{Motion Encoder}

we construct a Transformer-based encoder architecture to extract motion features. This architecture ingests temporally sequenced SMPL-X \cite{smplx} parameters as input. In order to maintain dimensional consistency with the other two modalities, we need to pool the temporal dimension. Inspired by \cite{vit,actor}, an extra learnable token has been introduced. The final output feature $F_m \in \mathbb{R}^{B \times 1 \times C} $ as the motion state is holistically processed without sampling operations.

\subsection{Vision-Audio-Motion Multi-Modal Learning}

After obtaining data from the three modalities mentioned above, we use pooling methods to reduce the number of dimensions for vision and audio features, while keeping them consistent with motion features. Then, the three are input into the feature fusion module to train the model's multi-modal capability. The core of the feature fusion module is a fully connected layer. The loss function is based on the Pearson linear correlation coefficient (PLCC) between the MOSs provided by the dataset and the predicted scores of the model. The formula is as follows:
\begin{equation}
\mathcal{L}_{\text{PLCC}} = \frac{1}{8}[ \text{MSE}(\textbf{p}, \textbf{t}) + \text{MSE}(\operatorname{Cov}(\mathbf{p}, \mathbf{t})\cdot\textbf{p}, \textbf{t})]
\end{equation}

where $\text{MSE}(\cdot, \cdot)$ stands for mean square error and $\textbf{p}$, $\textbf{t}$ denotes normalized predicted numerical scores and normalized MOSs, respectively.

\section{Experimental}

\subsection{Experimental Settings}

The Ges-QAer model is implemented with PyTorch framework and trained on two NVIDIA 3090 cards. The learning rate is 1e-4. Warm up and linear learning rate decay scheduler is used. The batch size is set to 10 and the number of training epochs is set to 20. All experiments for each method are re-trained on the Ges-QA dataset using 5-fold cross-validation. The reported performance of the Ges-QAer is evaluated on the final weights after training.

\subsection{Compared Methods}

Since no specific method has been proposed for evaluating A2G tasks, we select state-of-the-art methods from Audio-Video Quality Assessment (AVQA) and multi-modal alignment areas for comparison, including:
\begin{itemize}
\item AVQA: DNN-RNT \cite{anna1}, DNN-SND \cite{anna2}, and GeneralAVQA \cite{generalAVQA1,generalAVQA2}.
\item Multi-modal Alignment: AVID-CMA \cite{avid_cma}, VAST \cite{vast}, and ImageBind \cite{imagebind}. The latter two can also map text or other modal information to the same semantic space in addition to audio and video.
\end{itemize}

All methods were re-trained or fine-tuned on the Ges-QA dataset after loading default weights, with performance evaluated using Spearman rank-order correlation (SRCC), Pearson linear correlation (PLCC), Kendall rank-order correlation (KRCC), and root mean squared error (RMSE). As evidenced in Table.~\ref{tab1}, Ges-QAer outperforms all benchmarked approaches, demonstrating minimum improvements of 14.3\% and 12.1\% on the SRCC metric across the two dimensions, respectively. This enhancement is attributed to the integration of the motion modality and the strategic selection of corresponding encoders.

\subsection{Ablation Study}

We conducted ablation studies to validate the necessity of visual-audio-motion multi-modal learning. Specifically, while maintaining identical training hyper-parameters, we train Ges-QAer models with varying input configurations. The entries "\emph{w/o} motion encoder" and "\emph{w/o} video encoder" in Table.~\ref{tab1} demonstrate the performance of Ges-QAer under modality-deprived conditions.

Crucially, the removal of motion modality reduces the model to conventional visual-audio paradigms. Experimental results confirm that the use of all three modalities enhances comprehensive performance. In contrast, when prioritizing specific motion attributes while discarding video information, significant performance degradation occurs. This decline comes from the abandonment of substantial benefits offered by established video quality assessment methodologies.

\section{Conlusion}

In this paper, we construct Ges-QA, the first quality assessment dataset for A2G task. This dataset provides multidimensional Mean Opinion Score (MOS) for 1,400 samples and offers preliminary insights into emotional congruence issues in A2G generation. We propose Ges-QAer, a novel multi-modal learning method for multidimensional quality assessment for A2G content. Ges-QAer achieves state-of-the-art performance on our benchmark, demonstrating the potential to enhance user audio-visual experiences and improve output quality of A2G methods.

\clearpage
\clearpage
\bibliographystyle{IEEEtran}
\bibliography{IEEEabrv,refer}
\end{document}